\documentclass[a4paper,11pt]{article}
\pdfoutput=1 % if your are submitting a pdflatex (i.e. if you have
\usepackage{jinstpub} % for details on the use of the package, please
\title{Application of large area SiPMs for the readout of a plastic scintillator based timing detector}

\author[a]{C.~Betancourt,}
\author[b]{A.~Blondel,}
\author[a]{R.~Brundler,}
\author[a]{A.~D\"{a}twyler,}
\author[b]{Y.~Favre,}
\author[c]{D.~Gascon,}
\author[c]{S.~Gomez,}
\author[b,1]{A.~Korzenev%
\note{Corresponding author.},}
\author[b]{P.~Mermod,}
\author[b]{E.~Noah,}
\author[a]{N.~Serra,}
\author[b]{D.Sgalaberna,}
\author[a]{B.~Storaci}

\affiliation[a]{Physik-Institut, Universit\"{a}t Z\"{u}rich, Z\"{u}rich, Switzerland}
\affiliation[b]{DPNC, Universit\'{e} de Gen\`{e}ve, Gen\`{e}ve, Switzerland}
\affiliation[c]{Institut de Ci\`{e}ncies del Cosmos, Universitat de Barcelona, Barcelona, Spain}
\emailAdd{alexander.korzenev@cern.ch}

\abstract{In this study an array of eight 6 mm $\times$ 6~mm area SiPMs was coupled to the end of a long plastic scintillator counter which was exposed to a 2.5~GeV/$c$ muon beam at the CERN PS. Timing characteristics of bars with dimensions \mbox{150 cm $\times$ 6 cm $\times$ 1~cm} and \mbox{120 cm $\times$ 11 cm $\times$ 2.5~cm}  have been studied. An 8-channel SiPM anode readout ASIC (MUSIC R1) based on a novel low input impedance current conveyor has been used to read out and amplify SiPMs independently and sum the signals at the end. Prospects for applications in large-scale particle physics detectors with timing resolution below 100~ps are provided in light of the results.
}


\begin{document}

\maketitle
\flushbottom

%%%%%%%%%%%%% Introduction %%%%%%%%%%%%%%%%%%%%%%%%%%%%%%%%%%%%%%%%

\section{Introduction}
\label{sec:intro}

Silicon photomultipliers (SiPMs) are widely employed in high energy physics experiments. High photon detection efficiency and typical peak spectral sensitivity ranging between 400  and 500~nm make them very convenient for detecting photons produced in de-excitation processes evoked by ionizing particles in a plastic scintillator. Compactness and robustness of SiPMs make mechanical implementation easy. Furthermore, when assembled in a 2D array they can cover a sizable area, therefore can be considered as a relevant replacement for traditional photomultiplier tubes (PMTs). In particular, the sensors can be coupled directly to a scintillator bulk to provide a time resolution on a sub 100~ps level \cite{Bonesini:2014ira,Cattaneo:2014uya,Betancourt:2017kkz}.

The principal requirement for a precise time measurement is a short rise time of the signal. A large SiPM capacitance increases the rise time and width of the signal and worsens the time resolution. In this regard, a large monolithic sensor or many smaller sensors with common cathode and anode \cite{Bonesini:2014ira} are naturally limited in area. A reduction of the capacitance can be achieved by connecting SiPMs in series which decreases the rise time of the leading edge but also decreases the amplitude of a signal \cite{Cattaneo:2014uya,Cervi:2017kjz}. A parallel connection of sensors with an independent readout and amplification is another option to isolate the sensor capacitances from each other. In this case signals are summed up at the end. The scheme can be implemented either as a discrete circuit \cite{Aguilar:2016iit} or as an ASIC \cite{MUSIC}. The latter has a clear advantage of compactness and was adopted as an input stage of the acquisition system used in the study.

A typical time-of-flight (ToF) detector comprises an array of long bars covering a large surface which provides a fast trigger signal or can be used for particle identification. A clear advantage of a SiPM array is that it can take the form of the bar cross section, thus avoiding complex shape light-guides. Omitting light-guides in general reduces the dispersion of photons and decreases the price of a bar.

The test-bench used in this work can be considered as a prototype for the design of the timing detector of the SHiP experiment \cite{Anelli:2015pba} and  the ToF system proposed for the ND280/T2K upgrade \cite{Davide}. Furthermore, the test-bench is used as a test ground for a recently developed ASIC MUSIC\,R1 \cite{MUSIC} which was employed in a test-beam study for the first time.

This paper is organized as follows. Section \ref{sec:setup} describes the experimental set-up and devices under test. Section \ref{sec:results} presents the analysis procedure and results. A summary and outlook are then given in section \ref{sec:summary}.

%%%%%%%%%%%%%%%%% Experimental setup %%%%%%%%%%%%%%%%%%%%%%%%%%%%%%%%%%%

\begin{figure}[t]
\includegraphics[width=0.55\textwidth]{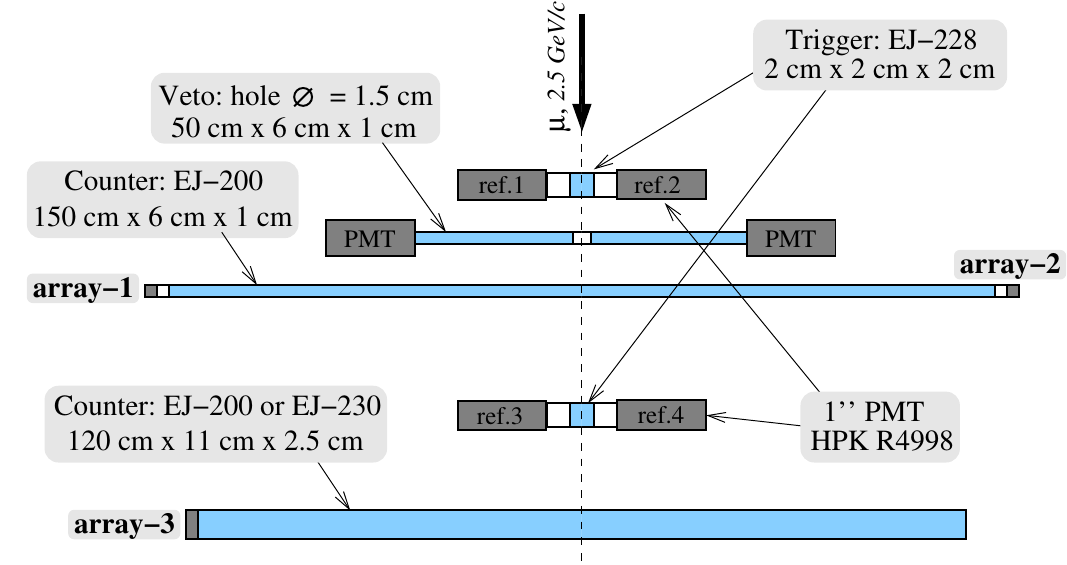}
\hfill
\includegraphics[width=0.42\textwidth]{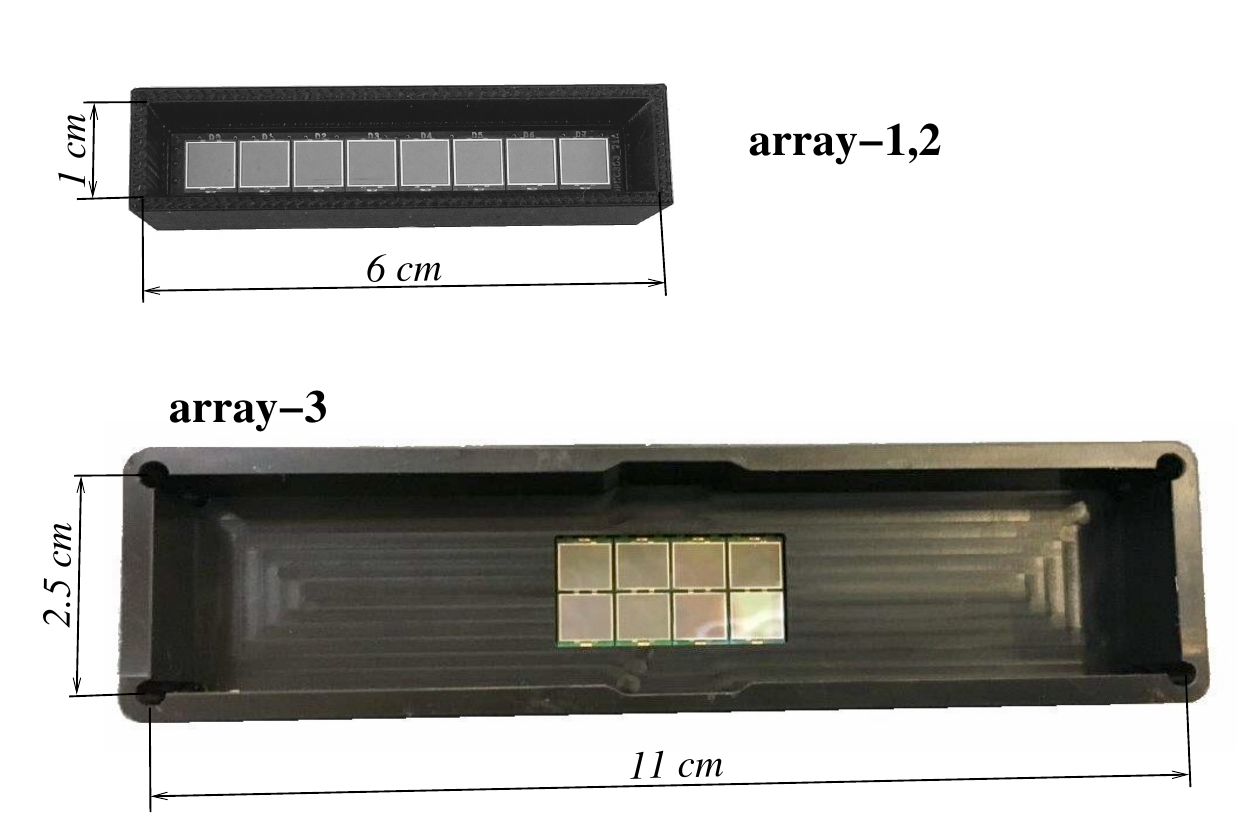}
\caption{ {\it Left}: schematic top view of the experimental setup.
  Outputs of all PMTs and SiPM-arrays were connected to a single acquisition module.
  {\it Right}: holders with the SiPM arrays.
}
\label{fig:setup}
\end{figure}

\section{Experimental set-up}
\label{sec:setup}

Measurements presented here were carried out in June of 2017 at the T9 beamline of the East Hall of the CERN PS. Three counters made of different materials and having different dimensions were exposed to a 2.5~GeV/$c$ muon beam to study their timing characteristics as a function of position. They are
\begin{itemize}

\item a 150 cm $\times$ 6 cm $\times$ 1 cm bar made of a EJ-200 cast plastic scintillator (attenuation length 380 cm, rise time 0.9~ns) \cite{SCIONIX}. Two 4~mm thick PMMA light-guides tapered from 60$\times$10~mm$^2$ to 56$\times$6~mm$^2$ area are glued to both ends of the bar.
%\footnote{A bar without light-guides (SiPMs directly coupled to the scintillator) was supposed to be studied as well. Unfortunately this bar has been damaged during transportation. In general, the effect of light-guides is not expected to be significant.}.

\item two 120 cm $\times$ 11 cm $\times$ 2.5 cm bars. One is made of a EJ-200 and another  made of a EJ-230. The latter has an attenuation length 120 cm and rise time 0.5~ns \cite{SCIONIX}. Light-guides have not been used.

\end{itemize}
The bars were wrapped in aluminum foil and black tape to ensure light tightness from the surrounding experimental hall. The scintillating light was read out by arrays of 8 SiPMs whose pulse shapes were recorded by a 16-channel waveform digitizer WAVECATCHER \cite{Delagnes:WAVECATCHER}. The digitizer was used at a 3.2~GS/s sampling rate. The circular buffer of WAVECATCHER contains 1024 cells allowing to cover a 320~ns time window, which records the full signal coming from SiPMs.

Eight surface-mount devices S13360-6050PE (area $6 \times 6$ mm$^2$, pixel pitch 50 $\mu$m) from Hamamatsu \cite{Hamamatsu} have been soldered in an array to a custom-made PCB, as shown in Fig.\,\ref{fig:setup}\,(right). Hereinafter the arrays are referred to as array-1 and array-2 for the 150 cm bar and array-3 for the 120 cm bars. Anode outputs of SiPMs have been read out and summed by an 8-channel SiPM anode readout ASIC (MUSIC R1) based on a novel low input impedance current conveyor  \cite{MUSIC}. Cathodes of all SiPMs are connected to a common power supply. The bias voltage of every SiPM was controlled via its anode using an internal DAC with 1\,V dynamic range. The ASIC circuit contains a tunable pole zero cancellation shaper which was used to reduce the SiPM recovery time constant. 
For the data taken with the 150 cm bar, a low gain configuration has been applied. 
Due to a larger cross section of the 120 cm bar, a smaller number of photons was expected
to be detected. Therefore the ASIC was used in a high gain configuration. 
In both cases the signals were summed in the chip after amplification.

The experimental setup is shown in Fig.\,\ref{fig:setup}\,(left). The trigger was formed by the coincidence of signals from two beam counters installed 20 cm up- and downstream of the beam with respect to the bar under test. Both counters used traditional PMTs for readout. The mean value of the times registered by both beam counters was considered as a reference and was subtracted from measurements of the main bar. A veto counter with a beam hole of 1.5 cm diameter was installed right after the first beam counter and was used in an anti-coincidence mode. The time resolution of the trigger system was found to be 20 ps.

The counters under test have been moved transversely with respect to the beam to study their time resolution as a function of the position of a charged-particle interaction along the the main axis of the bar. Hereinafter this axis referred to as $x$.

%%%%%%%%%%%%%% Analysis %%%%%%%%%%%%%%%%%%%%%%%%%%%%%%%%%%%%%%%%%%%%%%%%

\begin{figure}[t]
\centering
\includegraphics[width=0.4\textwidth]{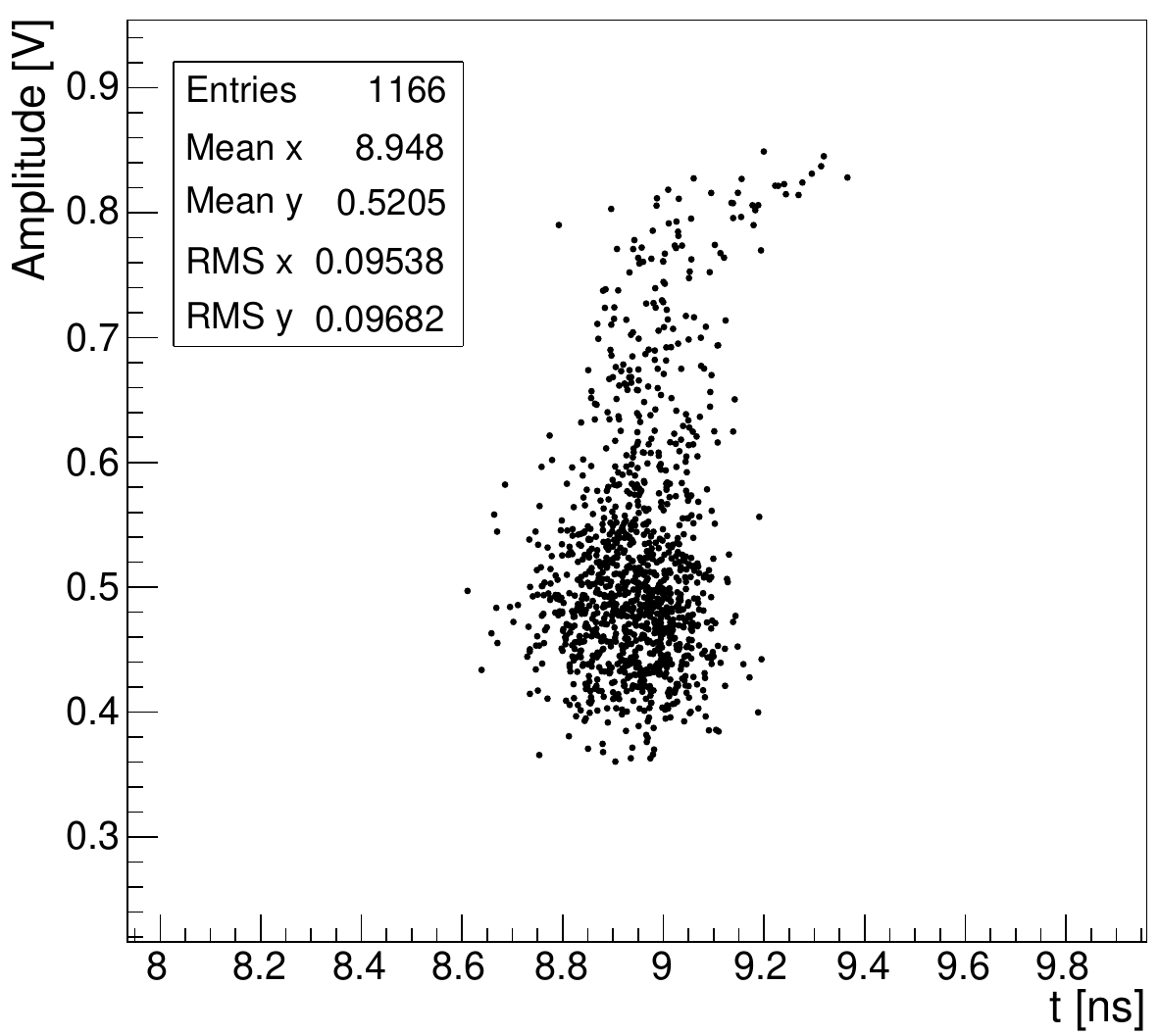}
\hspace{1.5cm}
\includegraphics[width=0.4\textwidth]{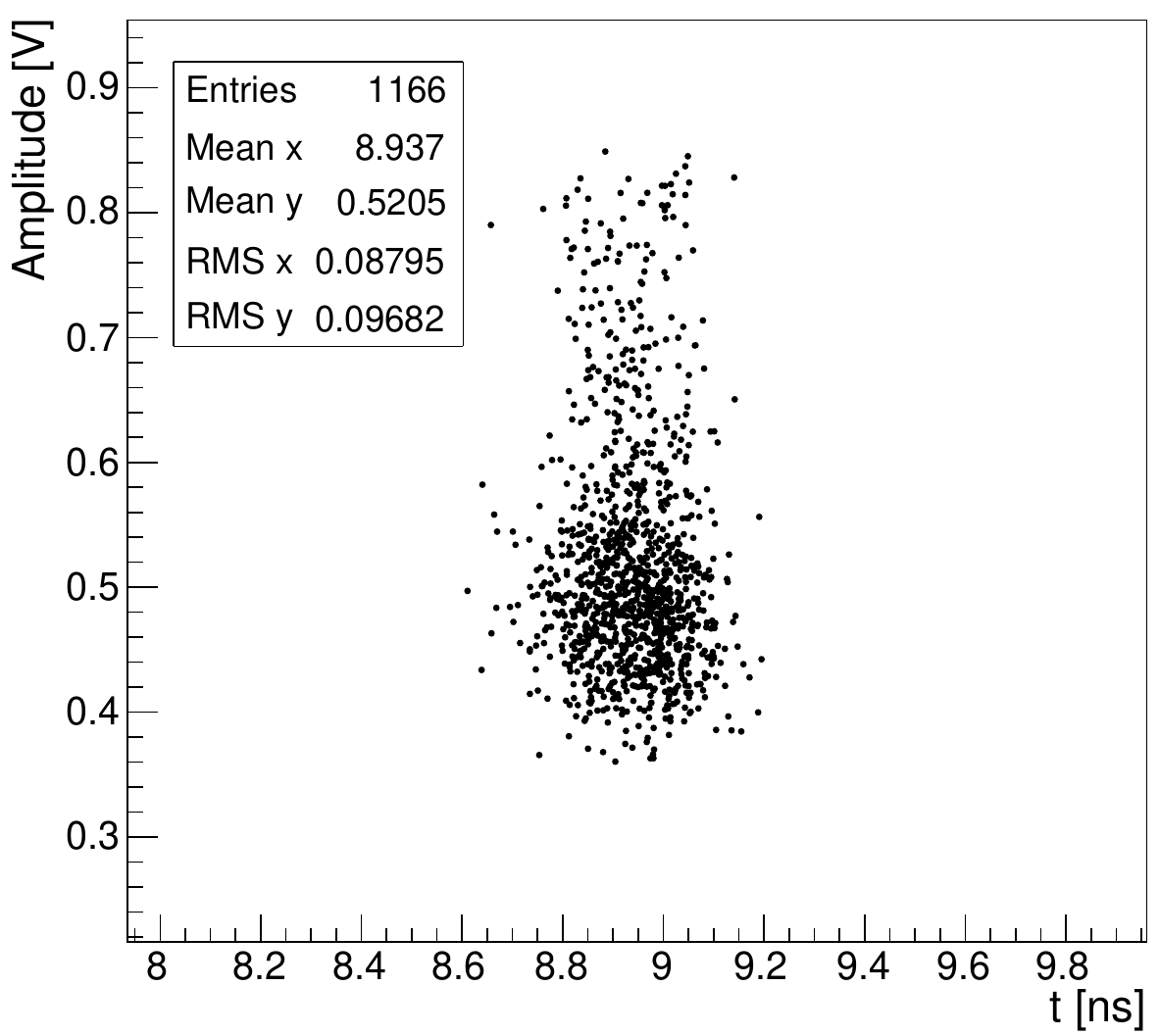}
\caption{ The correlation of amplitude versus time before ({\it left})
  and after ({\it right}) corrections for the measurement done with
  an interaction point located 10~cm from the array-2.
}
\label{fig:walk}
\end{figure}

\section{Analysis and results}
\label{sec:results}

Time responses of the SiPM arrays were calculated from registered waveforms in an offline analysis. A digital constant fraction discrimination (dCFD) technique was applied \cite{Delagnes:2016hdo}. In this approach the recorded waveforms were analyzed to find the signal height. An interval of the waveform before the signal was approximated by a constant to find the baseline. The time is defined by the crossing point of the interpolated digitized signal at the threshold which is a constant fraction of the pulse amplitude. 
%The threshold scan results in the 8\% fraction which provides an optimal value for the time resolution.
The results of the threshold scan defines the 8\% fraction as an optimal value for the time resolution.

It was found that for the values of amplitude larger than 0.6~V a correlation of the time versus amplitude appeared. An example of the correlation plot is shown in Fig.\,\ref{fig:walk}. The dependence of time on amplitude was parametrized using a polynomial function and corrections were applied in the analysis on the event-by-event basis. 
%This time-walk effect is significant for the interaction points located near to the bar ends where the number of detected photons is large.
Corrections for this time-walk effect improve the time resolution of the 150 cm bar by about 7\% for the interaction points located near the arrays where the number of detected photons is large. The effect is reduced to 4\% for the center of the bar.

%%%%%%%%%%%%%% Results %%%%%%%%%%%%%%%%%%%%%%%%%%%%%%%%%%%%%%%%%%%%%%%%%

\subsection{Results for the 150 cm $\times$ 6 cm $\times$ 1 cm bar}

\begin{figure}[t]
\centering
\includegraphics[width=0.482\textwidth]{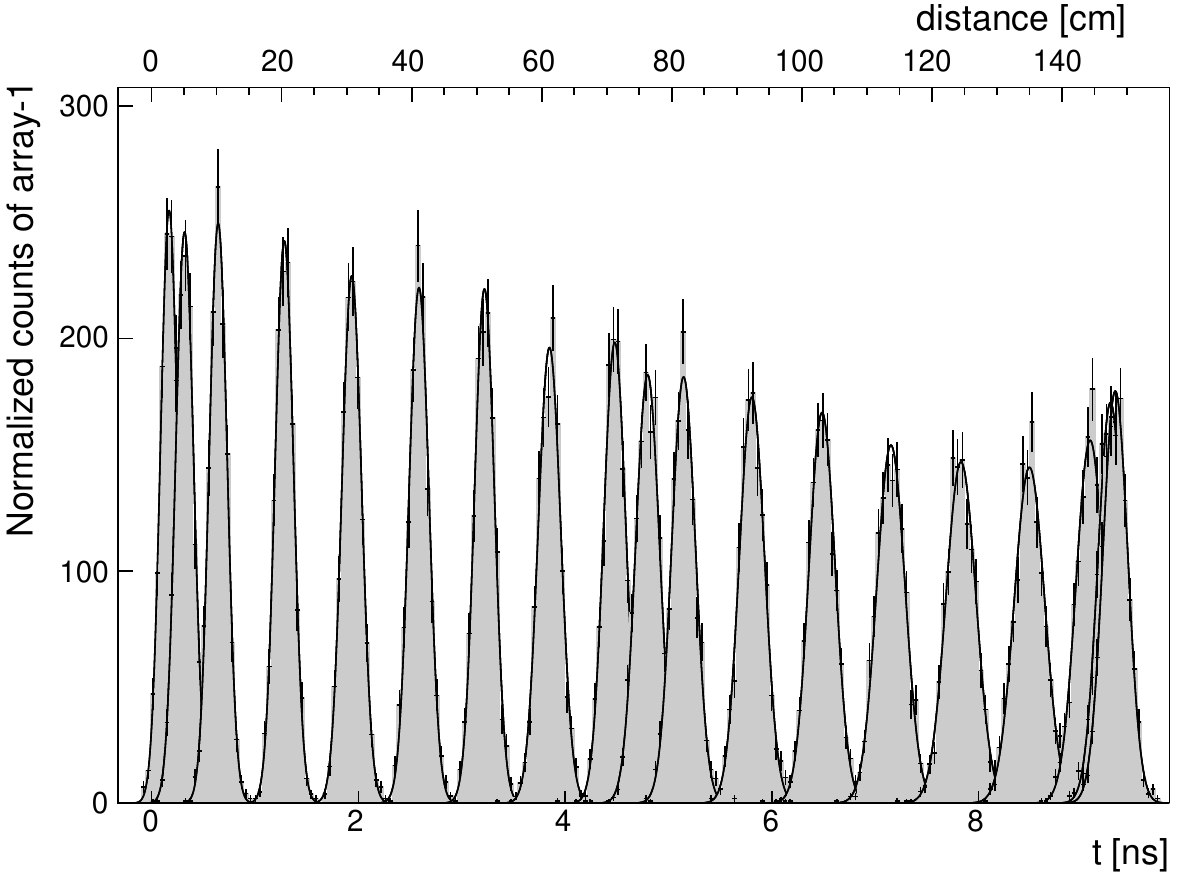}
\hfill
\includegraphics[width=0.50\textwidth]{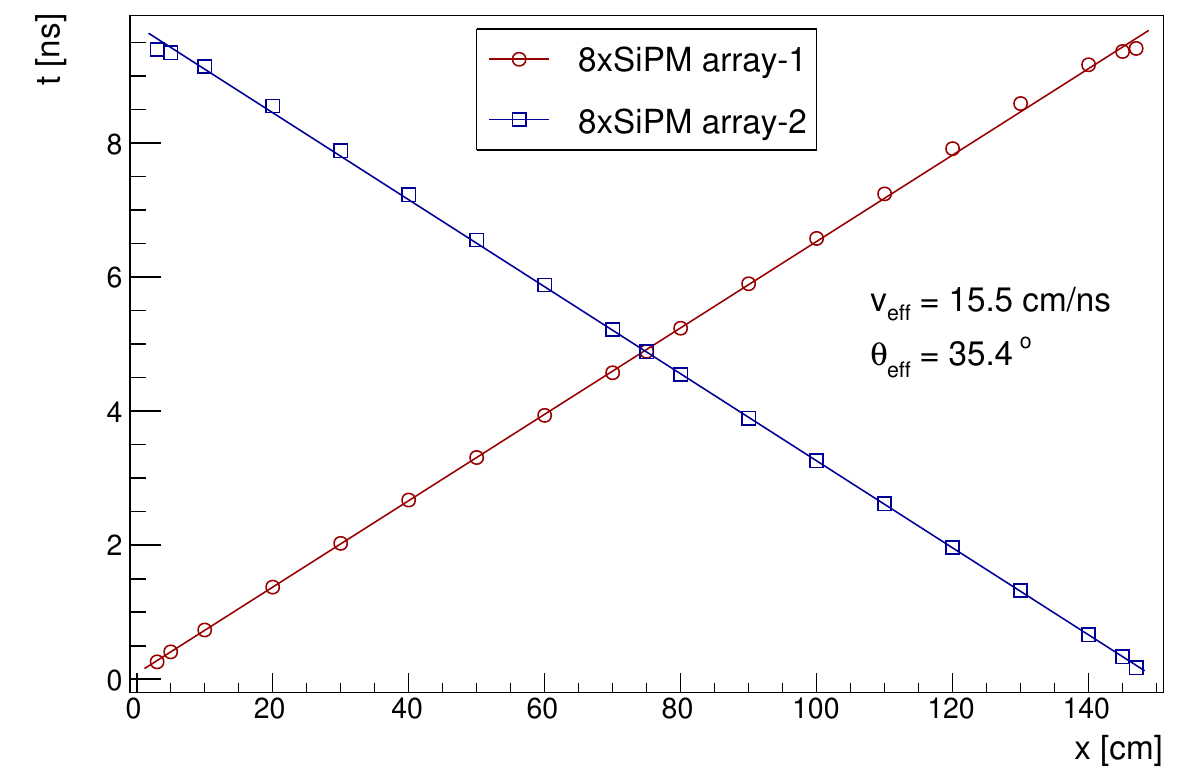}
\caption{Measurements done with the 150 cm $\times$ 6 cm $\times$ 1 cm bar.
  {\it Left}: superposition of distributions of time registered by the array-1 for
  different positions of an interaction point along the $x$ axis.
  Every distribution is approximated by a Gaussian function. 
  {\it Right}: the mean value of the function for both SiPM arrays is shown 
  as a function of the interaction position $x$ along the bar.
}
\label{fig:data}
\end{figure}

Examples of time spectra as measured for different positions along $x$ are shown in Fig.\,\ref{fig:data} (left). The time spectra can be reasonably approximated by Gaussian functions. For each position, the variance and the mean of the function were used to obtain the time resolution and the peak position of the distribution.

The dependence of the measured time versus position of the crossing point along the bar as viewed by both arrays is shown in Fig.\,\ref{fig:data} (right). The graphs are approximated by linear functions whose slopes represent the effective average speed of light along the $x$ axis, which is found to be $v_{eff} = 15.5$~cm/ns. One can convert this value into the effective average reflection angle using the refraction index of the plastic, which gives $\theta_{eff} = 35.4^\circ$.

\begin{figure}[t]
\centering
\includegraphics[width=0.49\textwidth]{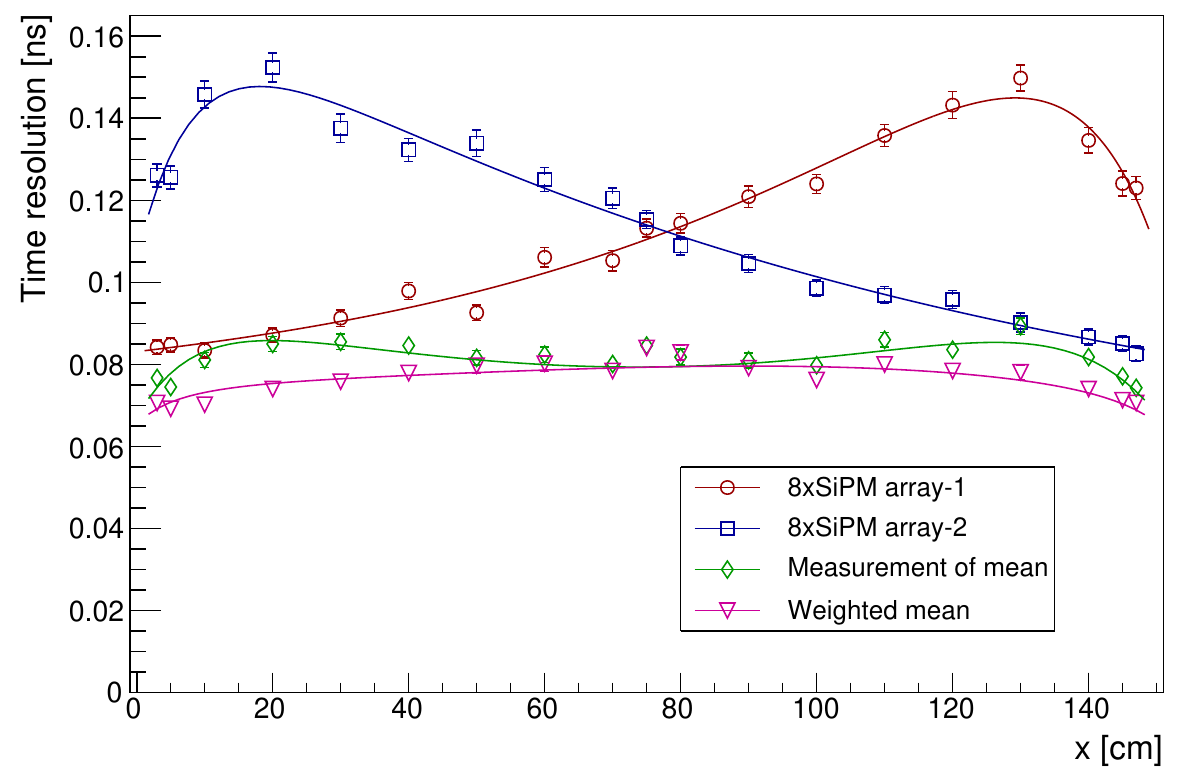}
\hfill
\includegraphics[width=0.49\textwidth]{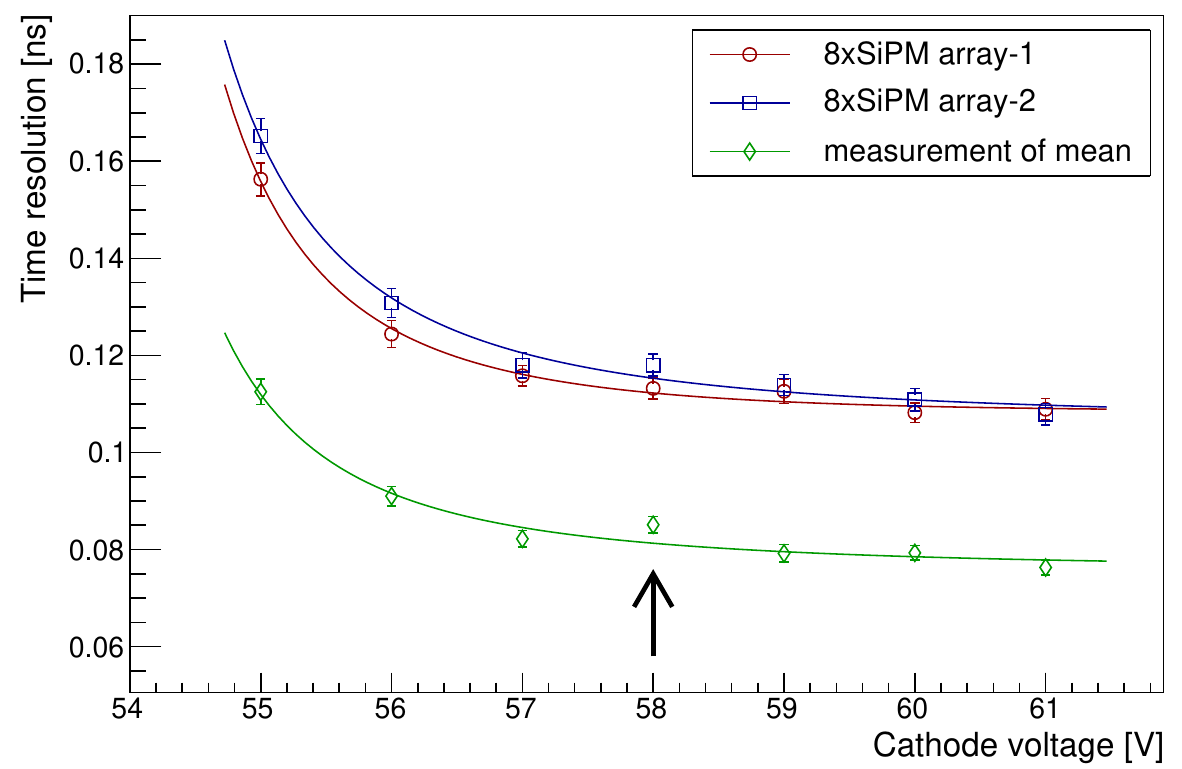}
\caption{Measurements done with the 150 cm $\times$ 6 cm $\times$ 1 cm bar.
  {\it Left}: time resolution as measured by the SiPM arrays at both ends
  of the bar as a function of the interaction point along the bar.
  {\it Right}: time resolution at the center of the bar ($x=75$ cm)
  as a function of the cathode voltage applied to the SiPM arrays.
  %The resolution corresponds to the mean time measurement shown in the left figure.
  %The voltage used in the data taking is indicated by an arrow.
  The voltage used for the data shown in the left plot is indicated by an arrow.
}
\label{fig:data_sigma}
\end{figure}

The time resolution of the counter as registered by the arrays is shown in Fig.\,\ref{fig:data_sigma} (left). It evolves from 83 ps for the crossing point near the sensor to 150~ps for the light propagation along the 130~cm distance. An improvement of the resolution is observed in case of the crossing point being at the proximity of $x=150$ cm. This could possibly be an effect of light reflected backwards. A similar effect was observed in Ref.\,\cite{Blondel:2016jju}. The distribution is approximated by an analytic function consisting of a sum of two exponential functions and a constant. The resolution  of the mean time and the weighted mean measurements is also shown. In both cases the time resolution is approximately 80~ps for the full length of the bar. However the weighted mean approach provides a visible advantage for interactions taking place in vicinity of the sensors.

The time resolution at the center of the bar as a function of a voltage applied to the common cathode of SiPMs is shown in Fig.\,\ref{fig:data_sigma} (right). One can clearly observe an improvement of the timing resolution with an increasing overvoltage. The improvement is prominent at lower voltage (close to the breakdown) and saturates for larger values. The 58\,V value, corresponding to about 5\,V overvoltage, was used for the measurements presented above.

%%%%%%%%%%%%%%%%%%%%%%%%%%%%%%%%%%%%%%%%%%%%%%%%%%%%%%%%%%%%%%%%%%%%%%%%

\subsection{Results for the 120 cm $\times$ 11 cm $\times$ 2.5 cm bar}

\begin{figure}[t]
\centering
\includegraphics[width=0.49\textwidth]{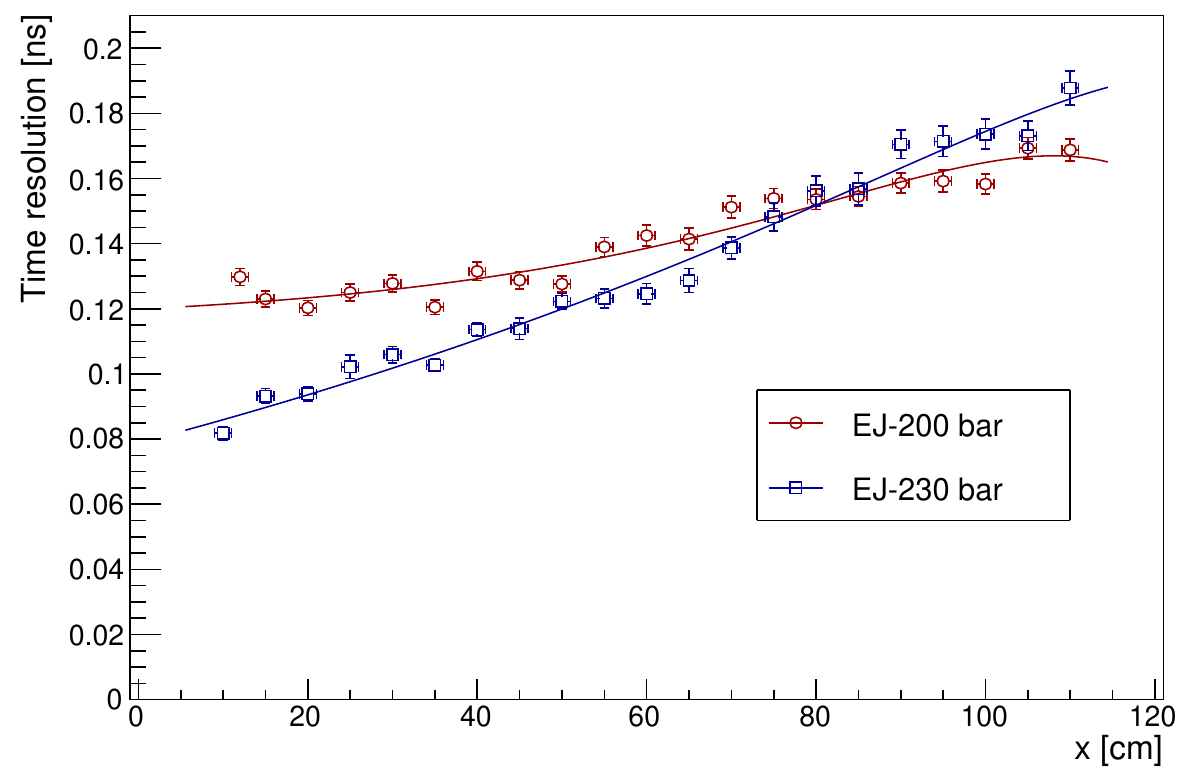}
\hfill
\includegraphics[width=0.49\textwidth]{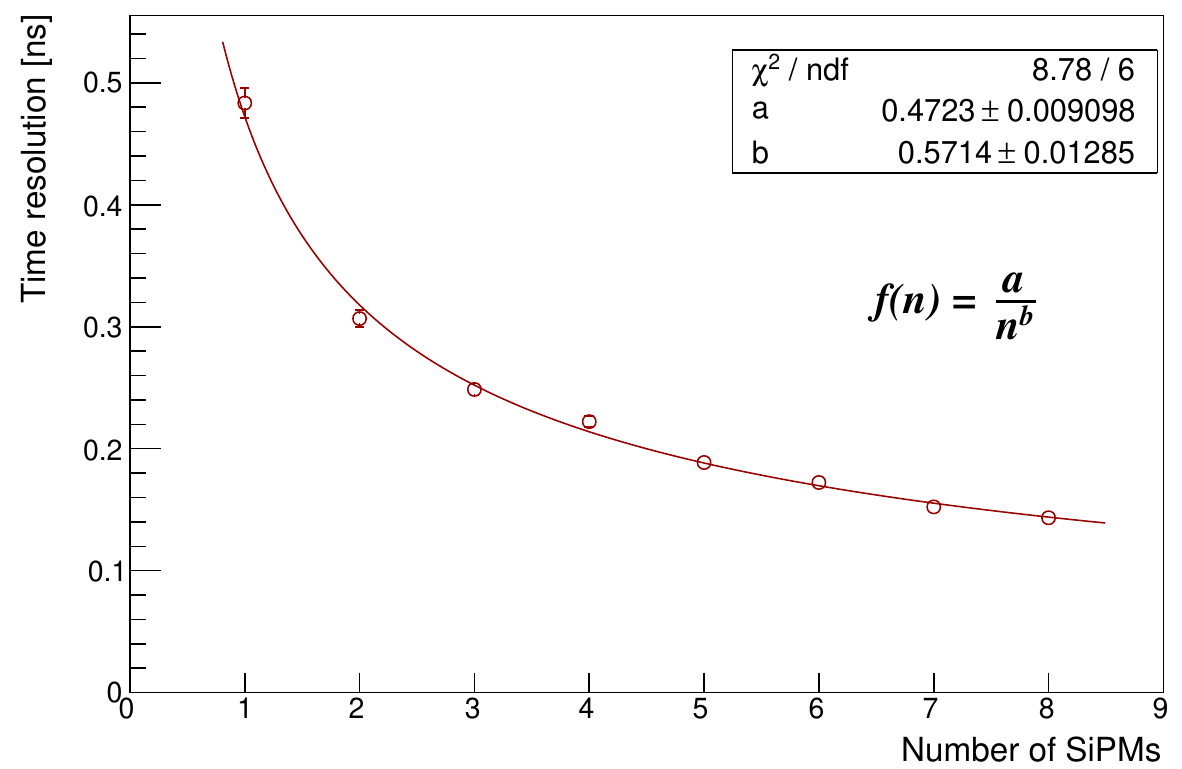}
\caption{Measurements done with the 120 cm $\times$ 11 cm $\times$ 2.5 cm bar.
  {\it Left}: time resolution as measured by the arrays-3 at one end
  of the bar as a function of the interaction point along the bar.
  Results are presented for EJ-200 and EJ-230 scintillators.
  {\it Right}: time resolution as a function of the number of SiPMs
  used in the readout. The measurement is done for interactions in
  the center of the EJ-200 bar.
}
\label{fig:UniZu}
\end{figure}

A similar analysis was performed for the bar with dimensions \mbox{120 cm $\times$ 11 cm $\times$ 2.5 cm}. Two scintillator materials, EJ-200 and EJ-230, were considered. Results for the time resolution as a function of distance are shown in Fig.\,\ref{fig:UniZu}\,(left).

The distribution for the EJ-200 bar can be compared to the results from the previous section presented in Fig.\,\ref{fig:data_sigma}\,(left). Since the SiPM sensitive area in both cases is the same the number of detected photons scales with the bar cross section area. This results in a simple ratio of bar widths; one takes a square root to convert this value to the ratio of the time resolutions giving $\sqrt{11~{\rm cm}/6~{\rm cm}} = 1.35$. Indeed, the data follow reasonably this prediction.

In general, the use of a large cross section bar together with a small sensor area is unpractical because of the light loss. However these results can be interesting in view of the replacement of a PMT readout in old experiments where scintillator counters already exist.

The EJ-230 material is used for very-fast timing applications. Its attenuation length is shorter as compared to EJ-200 (120 cm vs 380 cm), while it has a faster time response and lower self-absorption losses in the UV region. These properties basically define the behavior of the time resolution which is better for EJ-230 at small $x$ and worse for the far end of the bar.

Time resolution as a function of the number of SiPMs used in the readout (1--8) is shown in Fig.\,\ref{fig:UniZu}\,(right). The measurement was done for an interaction point in the center of the EJ-200 bar. The distribution of points can be reasonably described by the $1/\sqrt{n}$ behavior, where $n$ is the number of SiPMs in the readout chain.

%%%%%%%%%%%%%% Summary %%%%%%%%%%%%%%%%%%%%%%%%%%%%%%%%%%%%%%%%%%%%

\section{Summary}
\label{sec:summary}

A feasibility study of using an array of SiPMs for photon detection in a plastic scintillator has been presented. Compactness, mechanical robustness, high photon detection efficiency, low voltage operation and insensitivity to magnetic fields make SiPMs particularly useful for light collection in physics experiments. In this study two arrays of eight $6 \times 6$ mm$^2$ SiPMs have been coupled to both ends of a 150 cm $\times$ 6 cm $\times$ 1~cm plastic scintillator counter. Anode outputs of SiPMs have been read out and summed by an ASIC MUSIC\,R1. The time resolution as measured by a single array varies from 83~ps to 150~ps. The resolution in the case of the two sides readout is on average 80 ps. The resolution for the bar with a larger cross section is scaled according to the statistics of photons reaching the sensors. The technology has been proposed for the timing detector of the SHiP experiment at CERN SPS and the time-of-flight system of the T2K detector upgrade at JPARC.

%%%%%%%%%%%% Acknowledgment %%%%%%%%%%%%%%%%%%%%%%%%%%%%%%%%%%%%%%%%%%%

\acknowledgments

This work was supported by the Swiss National Science Foundation. We also would like to acknowledge the contribution of FAST (COST action TD1401) for inspiring a collaboration between the engineering group and the researchers. We thank the European Organization for Nuclear Research for support and hospitality and, in particular, the operating crews of the CERN PS accelerator and beamlines who made the measurements possible.


%\bibliographystyle{unsrtnat}
\bibliographystyle{JHEP}
\bibliography{jinst-SHiP_TD_2017}

\end{document}